\documentclass[conference]{IEEEtran}
\usepackage{cite}
\usepackage{amsmath,amssymb,amsfonts}
\usepackage{algorithmic}
\usepackage{graphicx}
\usepackage{textcomp}
\usepackage[table]{xcolor}

\usepackage[caption=false]{subfig}
\usepackage{multirow}
\usepackage{xtab}

\usepackage{color}
\usepackage{cite}

\usepackage{url}
\usepackage{breakurl}

\usepackage{hyperref}
\hypersetup{
   	colorlinks=true,       
	linkcolor=blue,        
	citecolor=blue,        
	filecolor=magenta,     
	urlcolor=black  
    }

\def\BibTeX{{\rm B\kern-.05em{\sc i\kern-.025em b}\kern-.08em
    T\kern-.1667em\lower.7ex\hbox{E}\kern-.125emX}}
\begin{document}

\title{A Preliminary Study on  the Role of Energy Storage and Load Rationing in Mitigating the Impact of the 2021 Texas Power Outage} 
\author{
\IEEEauthorblockN{Ali Menati and Le Xie}
\IEEEauthorblockA{\textit{Department of Electrical and Computer Engineering, and Texas A\&M Energy Institute} \\
\textit{Texas A\&M University}\\
College Station, Texas, United States \\
\text{\{menati,le.xie\}}@tamu.edu}
}
\maketitle 
\begin{abstract}
In February 2021, an unprecedented winter storm swept across the U.S., severely affecting the Texas power grid, leading to more than  4.5 million customers' electricity service interruption. This paper assesses the load shedding experienced by customers under realistic scenarios in the actual power grid. It also conducts a preliminary study on using energy storage and load rationing to mitigate rotating blackout's adverse impact on the grid. It is estimated that utility-scale battery storage systems with a total installed capacity of 920 GWh would be required to fully offset the load shedding during the Texas power outage if energy storage were the only technical option. Our simulation result suggests that implementing 20 percent load rationing on the system could potentially reduce this estimated energy storage capacity by 85 percent.  This estimate is obtained using the predicted capacity and demand profile from February 15 to February 18, 2021. Recognizing the fact that it would be very challenging to practically deploy energy storage of this size, approaches to provide more granular demand reduction are studied as a means of leveraging the energy storage to maximize the survivability of consumers. Preliminary case study suggests the potential of combining load rationing and proper sizing of energy storage would potentially provide much reliability improvement for the grid under such extreme weather conditions.  

\end{abstract}

\section{Introduction}
In recent years, energy storage capacity has witnessed a phenomenal growth in the U.S., reaching 1.6 GW of installed capacity in 2020 and tripling over the last five years \cite{EIAbattery}. With the fast-growing utility-scale energy storage installations, this number is expected to reach nearly 6 GW by the end of 2021 \cite{EIAbattery}. By looking into similar data from Texas, one can see that in the past, energy storage has not been attractive enough to thrive in ERCOT, but in recent years, it is witnessing a significant surge in the number of large-scale energy storage projects deployed for different purposes. As of August 2021, there is a total capacity of 459 MW of installed battery storage in the ERCOT grid \cite{ERCOTfact}. In this paper, a preliminary numerical simulation framework is developed to quantify the role of energy storage in mitigating the impact of power outages due to severe weather conditions. 

Researchers have been studying the ERCOT blackout of February 2021 from different perspectives. In \cite{Dongqi} the authors developed an open-source model to represent the Texas power grid, along with cross-domain data sets. They also evaluate multiple factors such as generation winterization to prevent power outages and compare their impact on the grid. In \cite{UTAustin} the 2021 power outage has been compared with similar events from the past, including the 1989 and 2011 blackout. They also analyze the economic consequences of the 2021 event.
The study in \cite{Profitability} shows that the loss of load in the February blackout was the most critical in the last 71 years. They argue that the large-scale investment in winterization of power generations is risky due to the low frequency of extreme cold-weather events. In this paper, a quantitative assessment of Texas blackouts is presented. Subsequently, energy storage and load rationing are proposed as two powerful mechanisms to prevent similar outages. We perform our simulations using the synthetic grid model developed in \cite{Dongqi}, along with the blackout-related data, including the generation and load-related data during the event period between February 15 to February 18. This is a 2,000-bus model that captures the key characteristics of the actual Texas electric grid, and it has been shown that the results reproduced by this model closely match the real-world load shedding done by ERCOT.

The value of Energy-Not-Served (ENS) is used to numerically describe the severity of the power outage event. ENS is a well-known index in the literature, calculated by taking the integral of load shedding over the blackout timeline. In this paper, a detailed assessment of load rationing is performed, but other voluntary approaches such as demand response are also capable of preventing load shedding. Demand response is a cost-effective scheme for demand peak reduction, which includes different methods and incentives used to increase the flexibility of the demand side. By convincing customers to move their power usage from the peak period to low demand times, both the grid and the customers greatly benefit from the resiliency and lower energy costs. In recent years, researchers have been focusing more on residential demand response, and the average peak reduction of 15 to 20 percent achieved by demand response has been shown extensively \cite{Coordination}.

In section~\ref{Schemes} energy storage and load rationing are introduced as two efficient approaches of mitigating power outages. Section~\ref{Storage} evaluates the Texas blackout event and quantifies the cost and capacity associated with implementing blackout prevention schemes. The potential impact of electric vehicle batteries on improving the grid's reliability is also analyzed in this section. The findings of our model-based simulations are presented in Section~\ref{simulation}, and different practical approaches are investigated to improve the survivability of most customers. Finally, section~\ref{conclusion} concludes the paper. 
\if{
\begin{figure}[t]
	\centering{}\includegraphics[ width=.82\columnwidth]{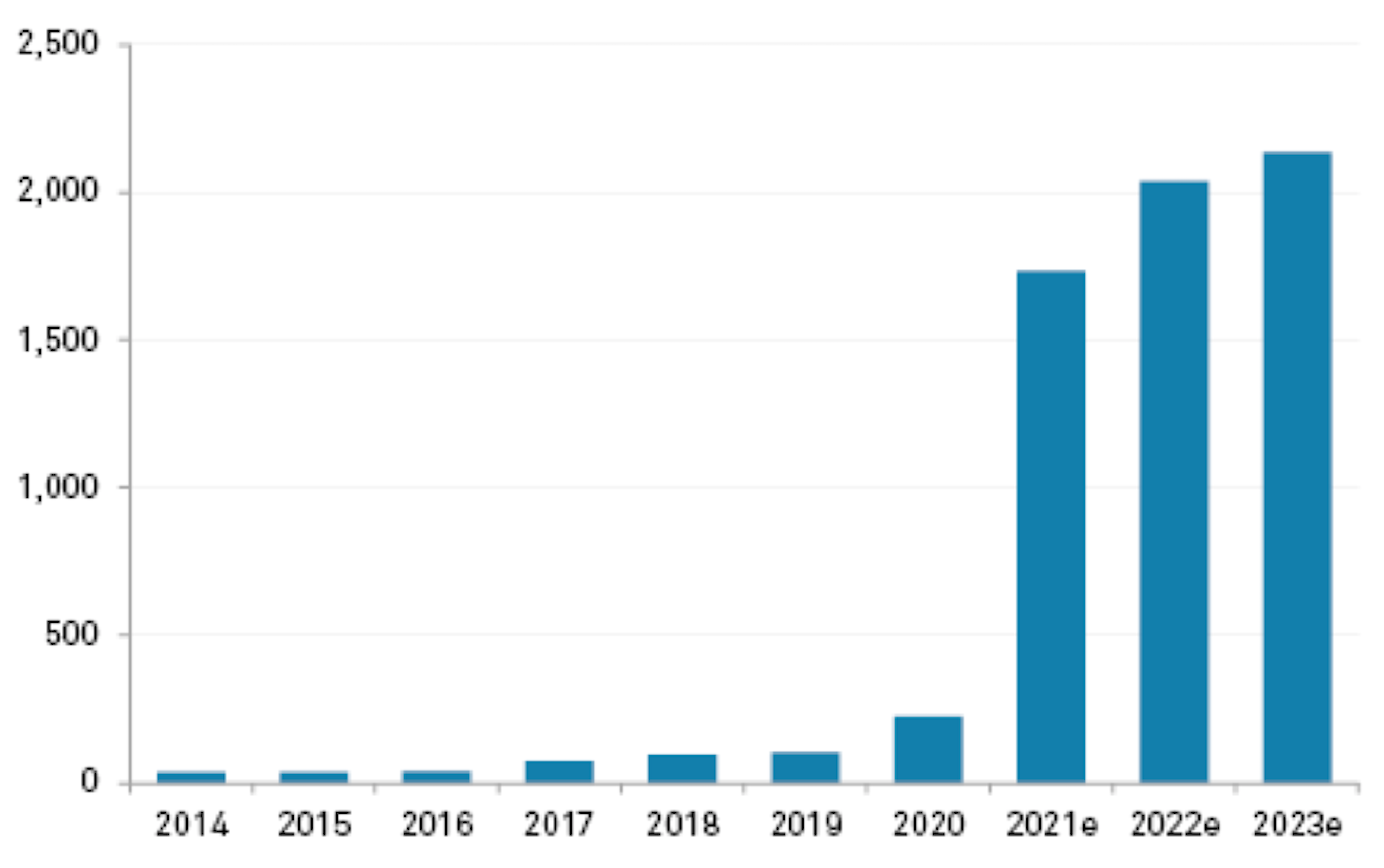}\caption{\label{capacity} Total installed and estimated cumulative ERCOT battery capacity (MW). Source: ERCOT.}
\end{figure}}
\fi

\section{Power Outage Mitigation Schemes}
\label{Schemes}
\subsection{Energy Storage}
In the past years, energy storage has played a relatively small role in Texas, but in the last year or two, the market has grown very quickly, and according to ERCOT, the total installed energy storage capacity is expected to rise above 2 GW by the end of 2022 \cite{sandp}. One of the primary incentives for energy storage companies in ERCOT is performing ancillary services. Although Texas has become very attractive for utility-scale storage providers, its total installed capacity is still a small fraction of ERCOT energy generation. In the following paragraphs, we explain some of Texas's prominent incentives and potentials for energy storage.

\begin{figure}[t]
	\centering{}\includegraphics[width=1\columnwidth]{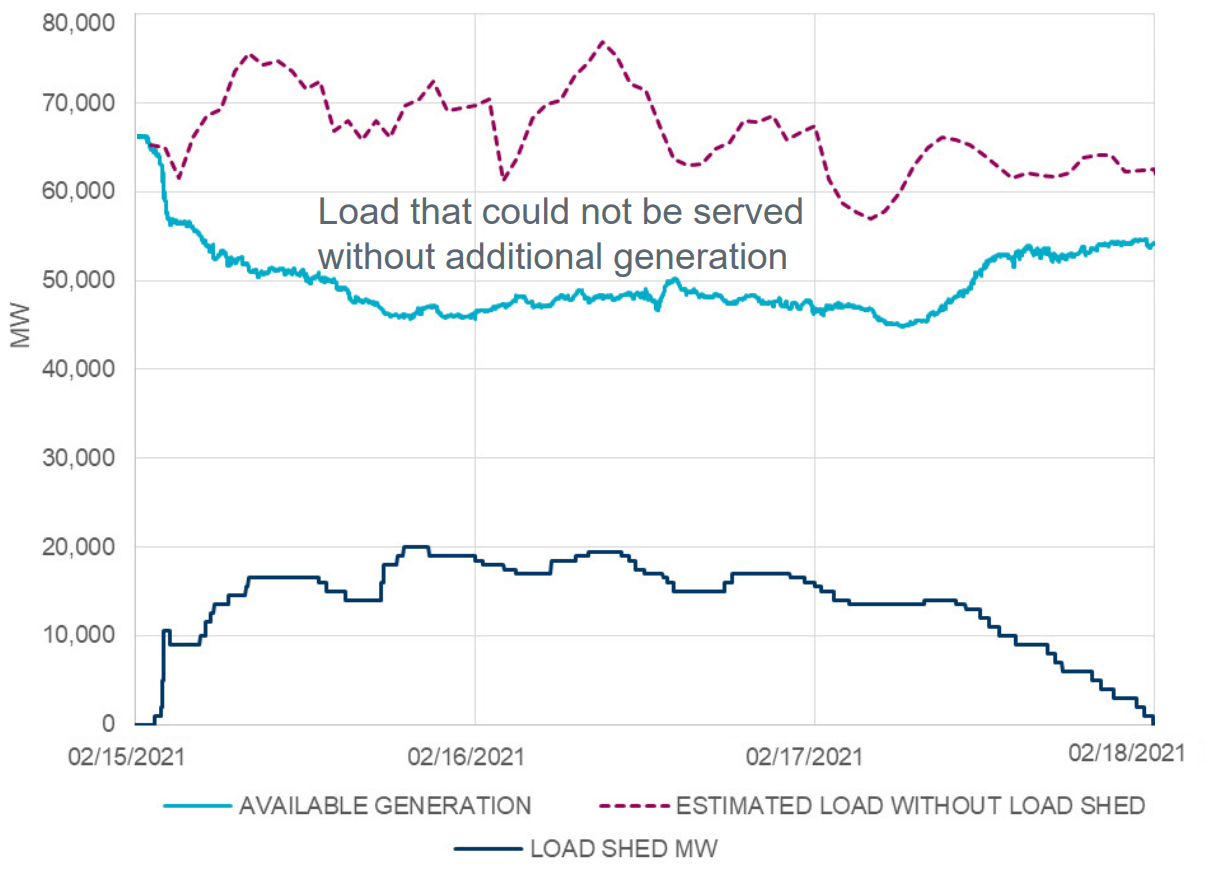}\caption{\label{fig1} ERCOT available generation capacity and its predicted load from February 15 to February 18. The real-world load shedding values demonstrate the widespread rotating blackouts. Source: \cite{ERCOTreview}.}
\end{figure}

\begin{itemize}
    \item Energy storage duration is a major factor in determining its price, and there are no duration requirements for batteries in ERCOT. This creates a unique opportunity where cheaper short-duration batteries are increasingly being used. The one or 2-hour batteries are more affordable than the advanced 4-hour batteries, and using them for ancillary services leads to lower costs and high revenues. In the past few years, we have witnessed a rapid increase in renewable energy generation capacity, including wind and solar. According to ERCOT, This trend will continue, and it is expected that in two years, from 2021 to 2023, around 35 GW of new wind and solar generation will come online \cite{sandp}. With the increasing renewable generation capacity and the growing solar and wind projects being deployed in Texas, the energy imbalance in the grid will also increase. Hence, power generation companies can use energy storage for online matching of supply to real-time demand.

    \item Another potential for energy storage is when the power generation sources are turned off and need to be restarted. This is known as the "black start" condition, where batteries can play a pivotal role in powering different parts of the grid. This provides a green and cost-effective solution for a quick restart of the power system.

\item Energy storage can also significantly improve and enable demand response programs used for peak reduction and increasing demand flexibility. In moving toward a highly renewable future grid, demand-side flexibility is imperative for balanced grid operation.

\end{itemize}
These unique opportunities are paving the way for the fast-growing energy storage market in ERCOT. In the long term, this storage capacity improves grid resiliency by providing energy reserves during extreme events. In Fig.~\ref{fig1} ERCOT's available online generation and its predicted load is depicted for the entire event timeline of February 15 to February 18. Due to the significant mismatch between generation and demand, load shedding continued for more than three days. We need to deploy load rationing alongside energy storage to reduce load shedding and maintain the grid's stability in similar events.

\subsection{Load Rationing}
Moving toward the future, long-duration extreme weather events happen more frequently. At the same time, with the increasing penetration of intermittent renewable energies into the grid, there is a growing need for an intelligent, flexible load management system that can deliver sufficient electricity to customers through a reliable system. Instead of using load shedding during difficult hours, our goal is to provide a minimum amount of electricity generation for each customer and cut off their electricity if they go beyond our limit. This is done through load rationing platforms, which can play a pivotal role in future load management systems by significantly reducing the economic and human cost of power outages. There are different ways to implement load shedding. For example, in \cite{automatic} a remote control is managing each customer's maximum power usage. If the users go beyond this limit, they receive a message notifying them about their cut-off within a short period. In \cite{loadrationing} a home load management system is proposed, in which the appliances with higher electricity consumption, such as washing machines, air conditioners, and heaters, are controlled through the designed platform that can switch on and off these devices during peak hours.

The residential sector makes up 33 percent of the total electricity demand in ERCOT \cite{nexus}, and with recent advances in the Internet of Things, cloud computing, and smart meters, online monitoring of customer usage is gaining more attention. By implementing an intelligent load rationing system along with energy storage, we can find the optimal combination that can entirely prevent power outages. In \cite{Idle} it is shown that the bare minimum of electric consumption is below $50$ percent. Hence, by providing the bare minimum of electricity consumption during extreme events, the total energy storage capacity needed to prevent power outages reduces significantly.

\section{Quantitative Assessment of ERCOT Blackout}
\label{Storage}

During the winter storm on February 15, 2021, due to the frigid temperatures across Texas, there was a sharp increase in demand and a severe drop in the generation capacity. Electricity demand in ERCOT reached 69 GW, and at the same time, nearly 30 GW of power generation was out. This high demand and shortage of power supply led to a significant mismatch between supply and demand, and in response, ERCOT ordered rotating outages across the state. Our simulating result using the synthetic grid shows that the Energy-Not-Served associated with this event is 920 GWh. Considering this number, preventing the power outage with energy storage as the only corrective measure seems impossible. The severity of this blackout can be seen in Fig.~\ref{fig1}, where the outages span over three days with the maximum load shedding of over 20 GW. In this section, the approximate cost of using energy storage to offset the blackouts is evaluated. Then we examine alternative methods capable of preventing similar events with lower costs.
\subsection{Implementation Cost}

One of the key reasons behind the recent surge in the worldwide deployment of energy storage is the fast reduction of its cost. The large-scale energy storage price has dropped $89\%$ rapidly decreasing from \$1,100/kWh to \$137/kWh \cite{BloombergNEF}. Considering this number, the cost of an energy storage system with a total installed capacity of 920 GWh, which can be used as the energy reserve for ERCOT, is
\begin{equation}
920 GWh \times \$137/kWh \approx \$126 \textit{B}.
\end{equation}
At the first look, this seems very large, but compared to the economic and human cost of the outages, it becomes more reasonable. It should be noted that currently, most of the energy storage in the grid, including the lithium-ion battery technologies, are short-term hourly batteries, which are not designed to survive long-term energy outages. 

While shorter duration energy storage can manage the intra-day fluctuation, Long Duration Energy Storage (LDES) is required for extreme events that last for multiple days. LDES helps achieve a net-zero carbon electricity system by shifting renewable energy. Considering global warming and the growing number of severe natural disasters, the need for backup power provided by LDES is increasing. As of today, the total installed energy storage in the world is near 173.7 GW. Despite the growing renewable penetration and the need for LDES, almost 93\% of this capacity is short-term storage \cite{DOE}. Energy arbitrage is an integral part of the financial advantages of any battery storage system, but even 4-hour energy storage can harness most of the arbitrage value, and increasing energy storage duration over a particular amount is not beneficial. Hence, considering the risk-averse nature of the customers in the utility industry, the wide-scale deployment of LDES is not practical at the moment. Since the total installed energy storage capacity in ERCOT cannot provide a large reserve, we need to propose practical approaches that could potentially close this gap. One of the most promising trends in this area is the electrification of the transportation section.

\subsection{EV Batteries in ERCOT}
\if{
\begin{figure}[t]
	\centering{}\includegraphics[ width=\columnwidth]{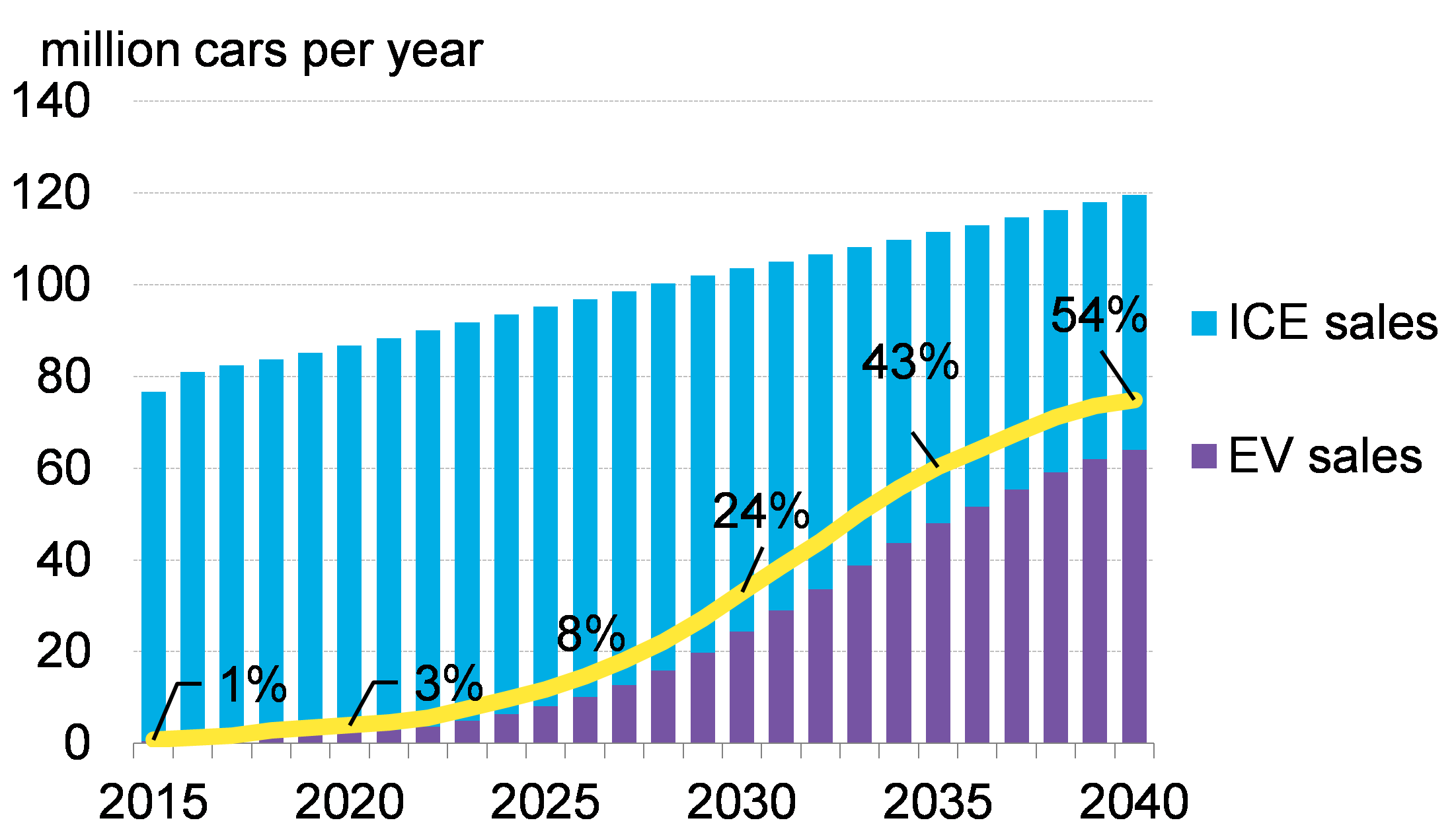}\caption{\label{cars}Long-term electric vehicle sales projection by BloombergNEF.}
\end{figure} 
}
\fi
\label{EV}

\begin{table}
\scriptsize
	\begin{center}
	
	\caption{}
		\label{EVpenetration}
	\centering
		\begin{tabular}{|c|p{1.4cm}|p{2.2cm}|p{1.9cm}|}
			\hline \rowcolor{-red} Types of Vehicle & Number of EVs in 2033& Per Vehicle Energy Capacity (kWh) & Total Energy Capacity (GWh)\\ 
			\hline  Cars &\,  3,000,000  & \, \, \,\, \, \, \, 20 & \, \, \,\, \, \,   60 \\ 
			\hline Short Haul/Buses  &\, \,  80,000 &\, \, \,\, \, \, \,350& \, \, \,\, \, \,   28 \\ 
			\hline  Long Haul Trucks & \, \,200,000  &\, \, \,\, \, \,\, 600 & \, \, \,\, \, \,120 \\ 
			\hline
		\end{tabular} 
	\end{center}
		
	EV penetration in Texas by 2033 \cite{Assessment}. Considering the number of vehicles and their corresponding battery capacity, they add 208 GWh of energy storage to ERCOT.
\end{table}

The widespread integration of electric vehicles into the grid could substantially transform the traditional power grid design and operation. It will increase the uncertainty of the grid and dramatically change the load patterns. Despite these challenges, if intelligently planned, EVs can increase the grid's resiliency by providing battery storage in the time of crisis. The global electric vehicle market size and adoption are expected to grow in the long run. Currently, only 3 percent of car sales worldwide are EVs, but this number is rapidly increasing, and it will reach 58 percent by 2040 when 33 percent of all the operational cars would be electric \cite{Bloomberg}. 

The regular size of an EV car battery is 60 kWh, which can provide enough energy for two days in a U.S. household. It should be noted that electric trucks and buses can hold ten times more energy than a regular EV \cite{EVcapacity}. According to ERCOT, there could be over 3 million electric vehicles in Texas by 2033 \cite{Assessment}. As shown in Table~\ref{EVpenetration}, this creates $60+28+120=208 \, \, GWh$ of potential energy storage, which provides an excellent opportunity for peak reduction and demand response. In the following section, we investigate the quantitative impact of implementing this energy storage along with residential load rationing.

\section{Numerical Simulation}
\label{simulation}
When the electricity reserve drops below a certain level in the Texas power grid and different resources such as demand response are insufficient to cover the mismatch between supply and demand, ERCOT issues Energy Emergency Alert (EEA) and asks transmission companies to shed load through rotating outages \cite{ERCOTEEA}. The size of load shedding depends on the predicted value of the online generation capacity and load on the grid. For our analytical purposes, load shedding is quantified as the gap between the predicted and post-shed load. In this paper, we use the post-shed electricity load data released by ERCOT with the 2,000-bus synthetic grid model developed in \cite{Dongqi} to simulate the events of the Texas Power outage in the period between February 15 to February 18. 

\subsection{Experiment Setting}
We assume that the load rationing in the system is only implemented in the residential sector, which is responsible for 1/3 of the total electricity demand in ERCOT \cite{nexus}. For example, by performing 60 percent load rationing on the residential sector, a 20 percent load reduction on the system is achieved. To obtain this percentage at difficult hours, the power consumption limit for each household is set to 40 percent of their normal electricity usage. As mentioned in section~\ref{Storage}, for severe long-duration weather events, current energy storage systems are not capable of preventing the power outage. But even for these events, the intelligent implementation of energy storage can effectively shape the load shedding curve and substantially reduce its peak. This paper follows this practical approach and measures the impact of the optimal combination of load rationing and energy storage in mitigating the power outage.  First, the load rationing scheme is developed to remove the bulk of load shedding and reduce it to a feasible scale that can be managed by battery storage. Then, energy storage reserve is applied to eliminate the peak of load shedding. Later in this section, we show that a 20 percent load rationing scheme reduces the total Energy-not-Served by more than 85 percent, and a regular-size 10 GWh energy storage reduces the peak of load shedding by more than 2 GW. 
\begin{figure}[t]
	\centering{}\includegraphics[ width=1\columnwidth]{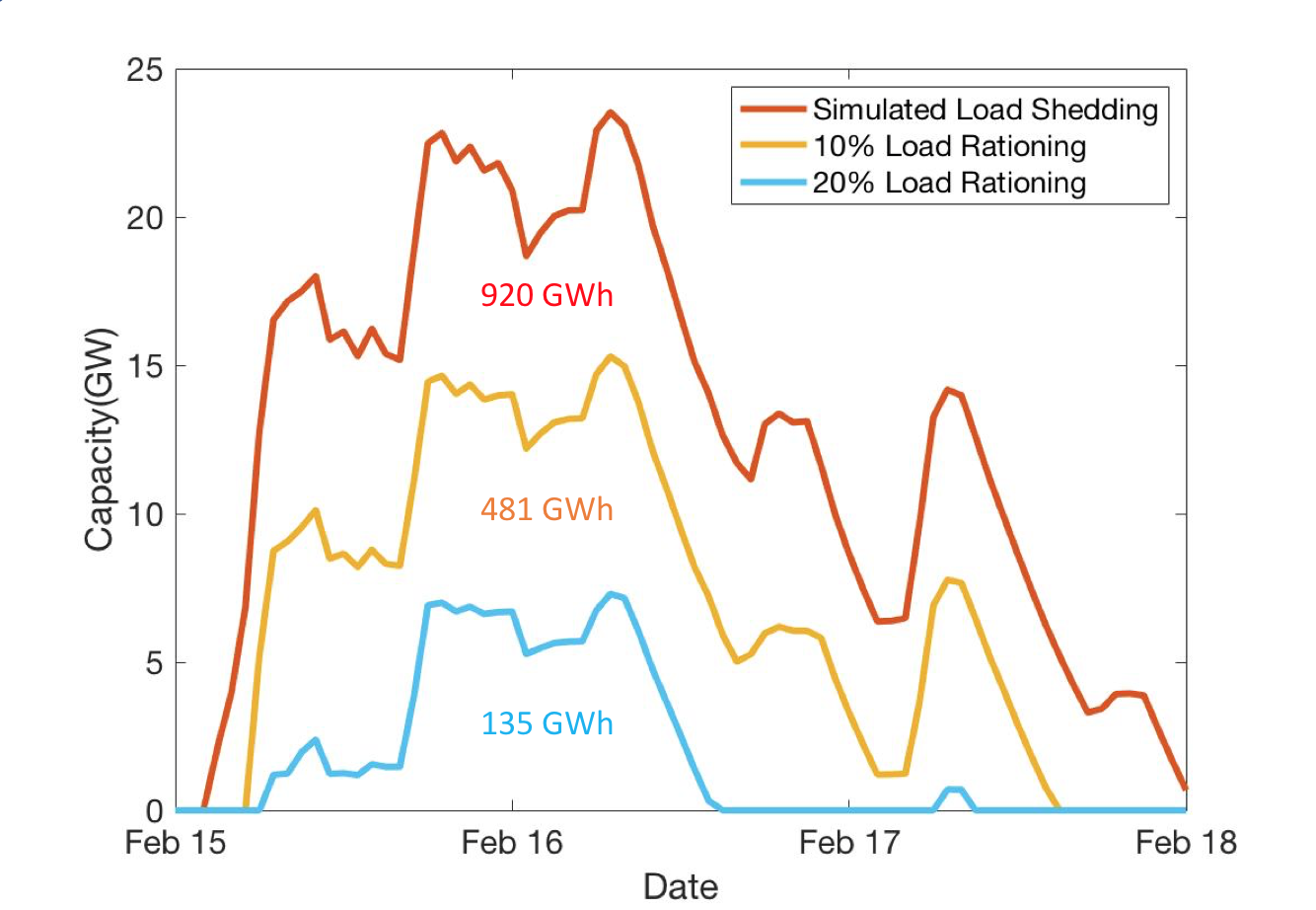}\caption{\label{loadshedding} Load shedding for different load rationing scenarios. Three possible designs with zero, 10, and 20 percent load rationing on the system are implemented, and their corresponding load shedding is presented in this figure. The simulated load shedding curve is a close approximation of the real-world load shedding done by ERCOT.
}
\end{figure} 
\subsection{Load Rationing Impact on Load Shedding}
In this section, the impact of load rationing on mitigating the blackout is examined for different scenarios. The simulated value of load shedding (GW) across the timeline is depicted in Fig.~\ref{loadshedding}. This curve represents the actual load shedding event, with no load rationing or energy storage deployed in the system. We also consider two cases of 10 and 20 percent load rationing with respect to the total load (30 and 60 percent for residential sector load, respectively).

As shown in Fig.~\ref{loadshedding}, with enough load rationing implemented on the system, avoiding the blackout becomes possible. However, it is imperative to ensure that people have access to the bare minimum of their electricity demand. Otherwise, without the reasonable upper limit on electricity consumption, load rationing fails to accomplish its desired objectives.
It can also be seen that load rationing flattens the curve, where due to its proportional nature, the peak of the load shedding reduces more than the valleys. Considering the total Energy-not-Served associated with each scenario, we notice that implementing 20 percent of load rationing substantially reduces the total ENS by more than 85 percent from 920 GWh to 135 GWh.

\subsection{Energy-not-Served for Different Scenarios}

\begin{figure}[t]
	\centering{}\includegraphics[ width=.985\columnwidth]{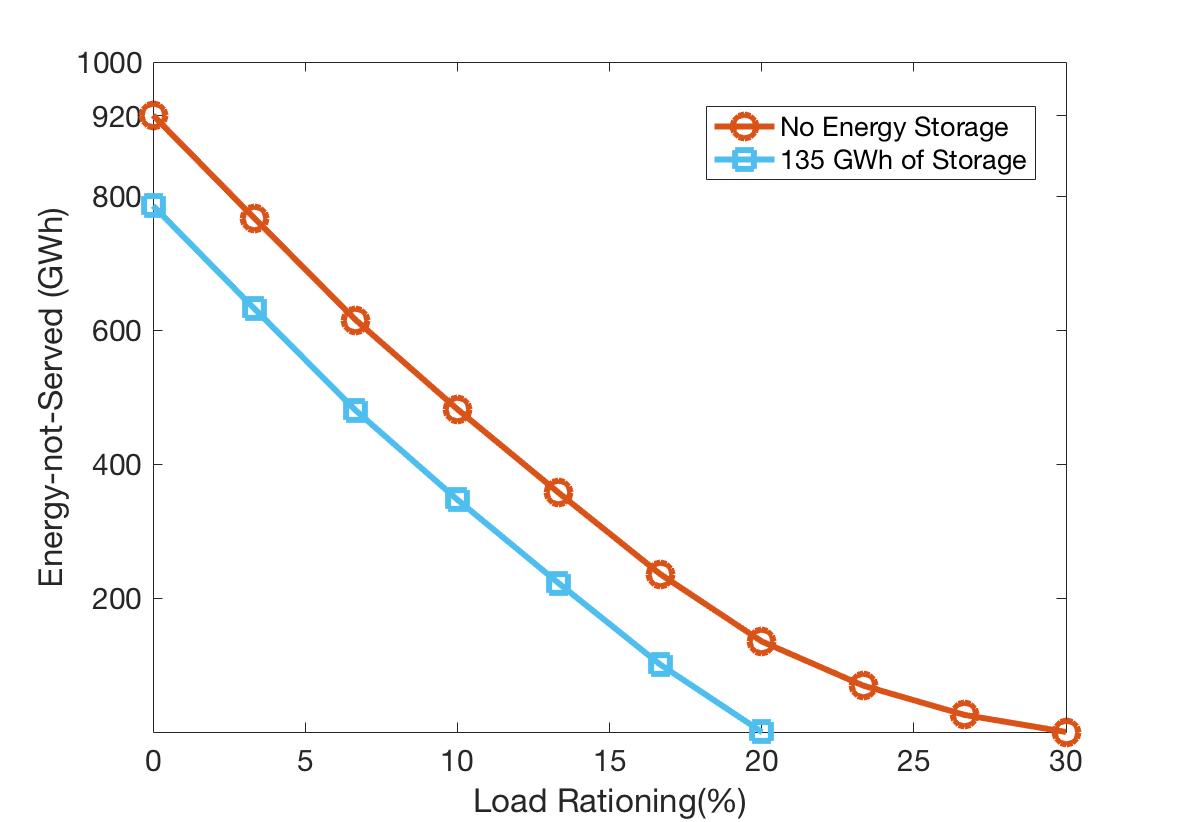}\caption{\label{ENS} The value of Energy-not-Served (ENS) for two energy storage availability scenarios in ERCOT. By implementing 135 GWh of energy storage, the load rationing percentage required for preventing the power outage significantly reduces. This shows the trade-off between the energy storage installation cost and customers access to electricity.}
\end{figure}

The value of Energy-not-Served as a function of load rationing percentage (with respect to the total load on the system) is depicted in Fig.~\ref{ENS}. We consider two scenarios, where in the first one, there is no energy storage connected to the grid, and in the second scenario, 135 GWh of storage is available as energy reserve. As one can see from this figure, 30 percent load rationing is needed to prevent the outage entirely when there is no energy storage. However, implementing 135 GWh of energy storage can prevent the load shedding using only 20 percent load raining. This observation suggests that although load rationing helps mitigate power outages, it can not entirely prevent them without deploying energy storage. 

Having 135 GWh of energy storage is not practical at the moment, and it is far from the current ERCOT storage potential, but it is not out of reach. As mentioned in section~\ref{EV}, by 2033, there would be over 208 GWh of EV batteries in ERCOT, which is beside the fast-growing utility-scale energy storage facilities being developed in Texas. As one can see in Fig.~\ref{ENS}, for the case with no energy storage, at first, the ENS reduces linearly with increasing the load rationing but, after 20 percent of load rationing, the reduction starts to slow down. This means that by implementing 20 percent load rationing, the value of ENS significantly declines by 85 percent, but to remove the remaining 15 percent, we need to further increase the load rationing to 30 percent. This is due to the shape of the load shedding curve, which has a wide base with a few peaks. By implementing load rationing, the flat part starts shrinking, but the peaks are relatively harder to remove, and we need to deploy energy storage for bringing the ENS down to zero.

\subsection{Load Shedding Peak Shaving}
\label{Peak-Clipping}
\begin{figure}[t]
	\centering{}\includegraphics[ width=1\columnwidth]{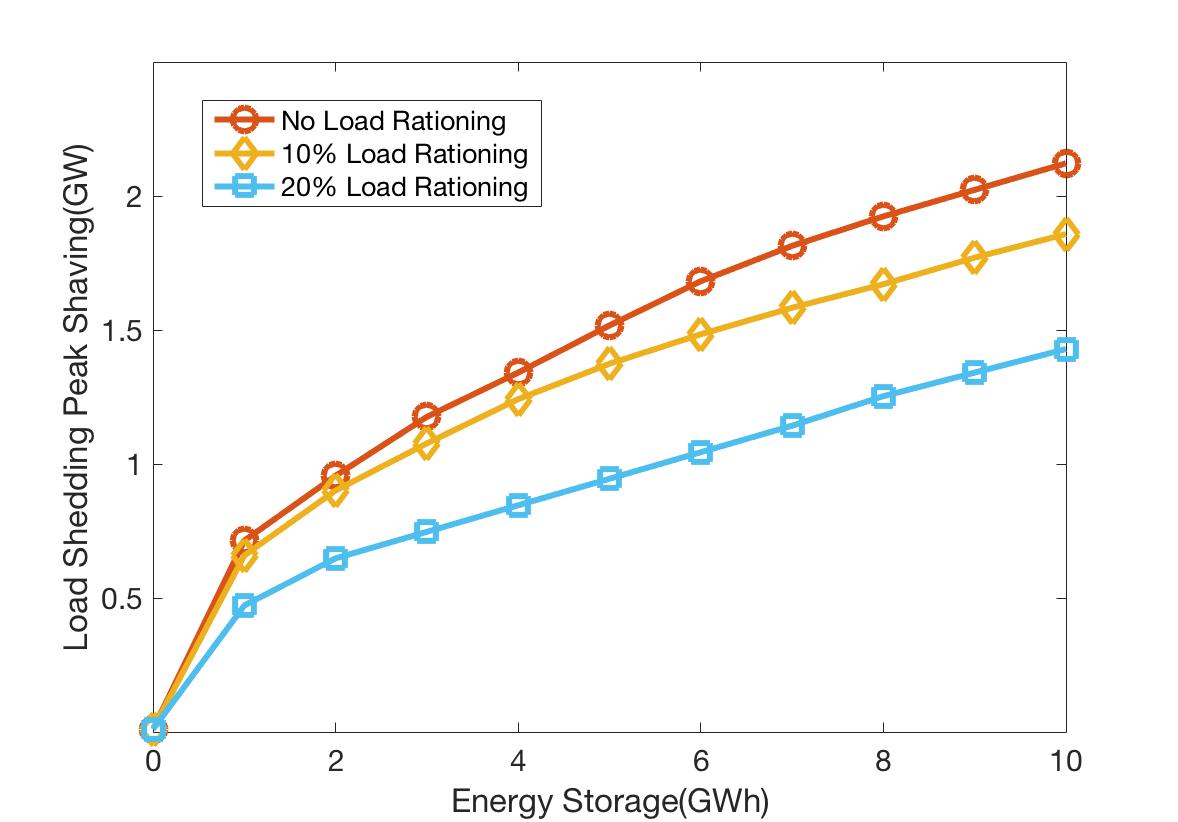}\caption{\label{peak} The reduction of peak load shedding (peak shaving) for different settings of load rationing. In this simulation, a realistic 10 GWh energy storage is deployed alongside three possible load shedding scenarios with zero, 10, and 20 percent load rationing.}
\end{figure}

In Fig.~\ref{peak}, the impact of energy storage on the system is investigated by assessing the reduction of peak load shedding (peak shaving) for different energy storage sizes. By adding more energy storage into the grid, the load shedding peak shaving increases, but the performance of per-GWh energy storage reduces. Here we consider three systems with different load rationing scenarios. By increasing load rationing on the system, the energy storage potential for peak shaving starts to decline. This is in line with our previous observation that load rationing flattens the curve of load shedding. It should be noted that in this simulation, load shedding peak shaving of more than 2 GW is achieved using less than 10 GWh of energy storage. 

\section{Concluding Remarks}
\label{conclusion}

In this paper, we conduct a preliminary study of February 2021 Texas blackouts and assess the role of energy storage and load rationing as two possible methods to mitigate the impact of severe weather events. Considering the current scale of energy storage availability in the system, more granular demand reduction approaches are investigated to maximize the survivability of most consumers. Building upon our proposed designs, we perform multiple simulations under realistic scenarios to examine the impact of load rationing and energy storage on the power outage severity. Our model-based simulation results suggest the potential of combining load rationing and proper sizing of energy storage would significantly improve the grid's reliability under extreme weather conditions. Future work will conduct more comprehensive investigation on the potential value of grid-edge technologies in mitigating the impact of severe weather on the power grid operation.  

\section*{Acknowledgement}
This work is supported in part by the National Science Foundation under Grant ECCS-2035688 and CMMI-2130945. Any opinions, findings, and conclusions or recommendations expressed in this material are those of the author(s) and do not necessarily reflect the views of the National Science Foundation.

\bibliographystyle{IEEEtran}

\bibliography{ref}
\end{document}